\documentclass[fleqn,usenatbib]{mnras}
\usepackage{graphicx}
\usepackage{newtxtext,newtxmath}

\usepackage[T1]{fontenc}

\DeclareRobustCommand{\VAN}[3]{#2}
\let\VANthebibliography\thebibliography
\def\thebibliography{\DeclareRobustCommand{\VAN}[3]{##3}\VANthebibliography}

\usepackage{graphicx}	
\usepackage{amsmath}	
\usepackage{subfigure}
\usepackage{float}
\usepackage{bm}
\usepackage{orcidlink}

\newcommand{\mpchi}{\,h^{-1}{\rm {Mpc}}}

\newcommand{\msun}{{\rm M}_{\sun}}
\newcommand{\msunh}{h^{-1}{\rm M}_{\sun}}

\title[Accurate halo catalogues with FastPM simulations]{Improving the Accuracy of Halo Mass Based Statistics For Fast Approximate N-body Simulations}

\author[Wu et al.]{%
Yiheng Wu$^{1,2,3}$\orcidlink{0009-0004-6083-6608},
Hong Guo$^{1}$\orcidlink{0000-0003-4936-8247}\thanks{E-mail: guohong@shao.ac.cn},
Volker Springel$^{3}$\orcidlink{0000-0001-5976-4599}\thanks{Email: vspringel@mpa-garching.mpg.de}
\\%
$^{1}$Shanghai Astronomical Observatory, Chinese Academy of Sciences, Shanghai 200030, China\\%
$^{2}$University of Chinese Academy of Sciences, Beijing 100049, China\\%
$^{3}$Max-Planck-Institut f\"{u}r Astrophysik, Karl-Schwarzschild-Stra\ss{}e 1, 85740 Garching bei M\"{u}nchen, Germany
}

\begin{document}
\label{firstpage}
\pagerange{\pageref{firstpage}--\pageref{lastpage}}
\maketitle

\begin{abstract}
Approximate N-body methods, such as FastPM and COLA, have been successful in modelling halo and galaxy clustering statistics, but their low resolution on small scales is a limitation for applications that require high precision. Full N-body simulations can provide better accuracy but are too computationally expensive for a quick exploration of cosmological parameters. This paper presents a method for correcting distinct haloes identified in fast N-body simulations, so that various halo statistics improve to a percent level accuracy. The scheme seeks to find empirical corrections to halo properties such that the virial mass is the same as that of a corresponding halo in a full N-body simulation. The modified outer density contour of the corrected halo is determined on the basis of the FastPM settings and the number of particles inside the halo. This method only changes some parameters of the halo finder, and does not require any extra CPU-cost. We demonstrate that the adjusted halo catalogues of FastPM simulations significantly improve the precision of halo mass-based statistics from redshifts $z=0.0$ to $1.0$, and that our calibration can be applied to different cosmologies without needing to be recalibrated.
\end{abstract}

\section{Introduction}

Numerical simulations are a key tool for cosmological studies to determine the fundamental characteristics of the universe \citep[see e.g.,][for a review]{Angulo2022}. These simulations can be used to generate mock catalogues of the real universe with varying cosmological parameters. Full N-body simulations with large volume and high mass resolution are still very expensive in terms of CPU time and disk storage \citep[e.g.][]{Potter2017, Heitmann2019, Ishiyama2021,Hernandez-Aguayo2023}. Approximate N-body solvers \citep{Colavincenzo2019,Lippich2019,Monaco2016} provide an alternative to full N-body simulations to quickly produce realisations of the large-scale structure (LSS). These approximate methods include low-resolution N-body techniques such as FastPM~\citep{Feng2016,Chartier2021} and COLA~\citep{Tassev2013,Howlett2015,Izard2016,Izard2018,Wright2023,Ding2023}, and schemes based solely on Lagrangian perturbation theory such as PINOCCHIO~\citep{Monaco2002,Rizzo2017}. More recent developments include the use of deep neural networks to attempt to predict the formation of nonlinear structures of the universe \citep[e.g.][]{He2019}.

Although the halo catalogues from approximate simulations are correlated to those from full N-body simulations with the same initial conditions, the dark matter distribution on small scales is substantially less accurate. Increasing the particle-mesh resolution and performing finer timesteps may enhance the accuracy on small scales, but the resulting improvement may not be worth the additional computational cost if the precision stays below the several percent level for the dark-matter power spectrum at scales $k>1\,h\,{\rm Mpc}^{-1}$. The reason for a deficit of the predicted matter power spectrum at these scales in the approximate methods is their inability to resolve the detailed density profile of haloes around the centre \citep[e.g.][]{Neto2007}, as well as the substructure they contain \citep[e.g.][]{Gao2012}. For the same reason, the haloes actually reach a modified virial equilibrium where the density profile is much shallower than the NFW profile \citep{Navarro1997} expected from high-resolution simulations, and this leads to an underestimate of the halo mass function compared to full N-body simulations. These defects have a negative effect on the utility of the approximate simulations for cosmological applications in studies of large-scale structure, weak gravitational lensing, SZ effect, etc.

Despite the advantages of approximate methods, there have been efforts to improve them while preserving their low computational cost. \citet{Dai2018,Dai2020} proposed a gradient-based approach that mimics the short-range force that is absent in FastPM simulations. After calibration, the matter power spectrum can be improved to a precision of a few percent for a range of wave numbers $k$ and different redshifts. However, the free parameters in this method must be calibrated for different simulation settings. Furthermore, their research focused on the properties and substructures of FOF haloes. \citet{Fiorini2021} modified the FOF haloes identified in COLA to spherical overdensity haloes to facilitate research on modified gravity models.

This paper presents a method for correcting the mass $M_{200c}$ of individual haloes, so that the accuracy of basic halo statistics is improved to the percent level when compared to full N-body simulations. Our approach is based on the fact that the particle distribution on small scales is less concentrated in approximate simulations, meaning that the halo radii contain fewer particles than in full N-body simulations. This results in the overdensity of the material belonging to the halo not reaching a density contrast of $200$ times the critical density, but a lower value. To quantify this effect, we ran FastPM simulations with different particle numbers and cosmological parameters to compare them with full N-body simulations. This enabled us to study the relation between the effective halo density and the number of particles in the haloes. This, in turn, allows us to apply a fiducial halo overdensity, resulting in a corrected halo mass estimate. An advantage of this method is that the free parameters of the correction method are only dependent on the starting redshift and the number of time steps used for the FastPM simulation and are not significantly influenced by other cosmological parameters.

This paper is organised as follows. We first describe in Section~\ref{section:method} the details of our correction method for approximate simulations and then investigate its performance for different halo statistics, including the halo mass function and the halo clustering power spectrum in real and redshift spaces. In Section~\ref{section:cosmology} we run a set of five pairs of full N-body and matching FastPM simulation boxes with different cosmological models to test the applicability of our method to different cosmologies. In Section~\ref{section:redshift}, we then extend the correction method to higher redshifts. In Section~\ref{section:HOD}, we apply our method to the creation of mock galaxy catalogues with halo occupation distribution modelling. Finally, we discuss our results in Section~\ref{section:discussion}, and  summarise in Section~\ref{section:Summary}.

\section{Methods}\label{section:method}

In this section, we provide a brief overview of the approximate FastPM approach and discuss the details of the full N-body simulations of the IllustrisTNG300-Dark series, which we use for comparison. We use the same initial conditions for the N-body simulation and run FastPM for a range of particle numbers to empirically determine the relationship between the halo density, the halo particle number, and the total particle number density.

\subsection{Simulation Input}

\subsubsection{Approximate FastPM method}

FastPM \citep{Feng2016} is a powerful and efficient approximate particle-mesh N-body solver that uses modified kick and drift operators to ensure that linear growth is accurately reproduced on large scales, even with a limited number of time steps. It has the same initial condition generator as the N-GenIC~\citep{Springel2015} and {\small AREPO} code~\citep{Springel2010, Weinberger2020}, meaning the same initial conditions are generated when using the same random number seed. Alternatively, FastPM can also be used with externally prescribed initial conditions, such as a linear density perturbation field at $z=0$ that is then scaled back to the starting redshift.

\begin{figure*}
\includegraphics[width=\textwidth]{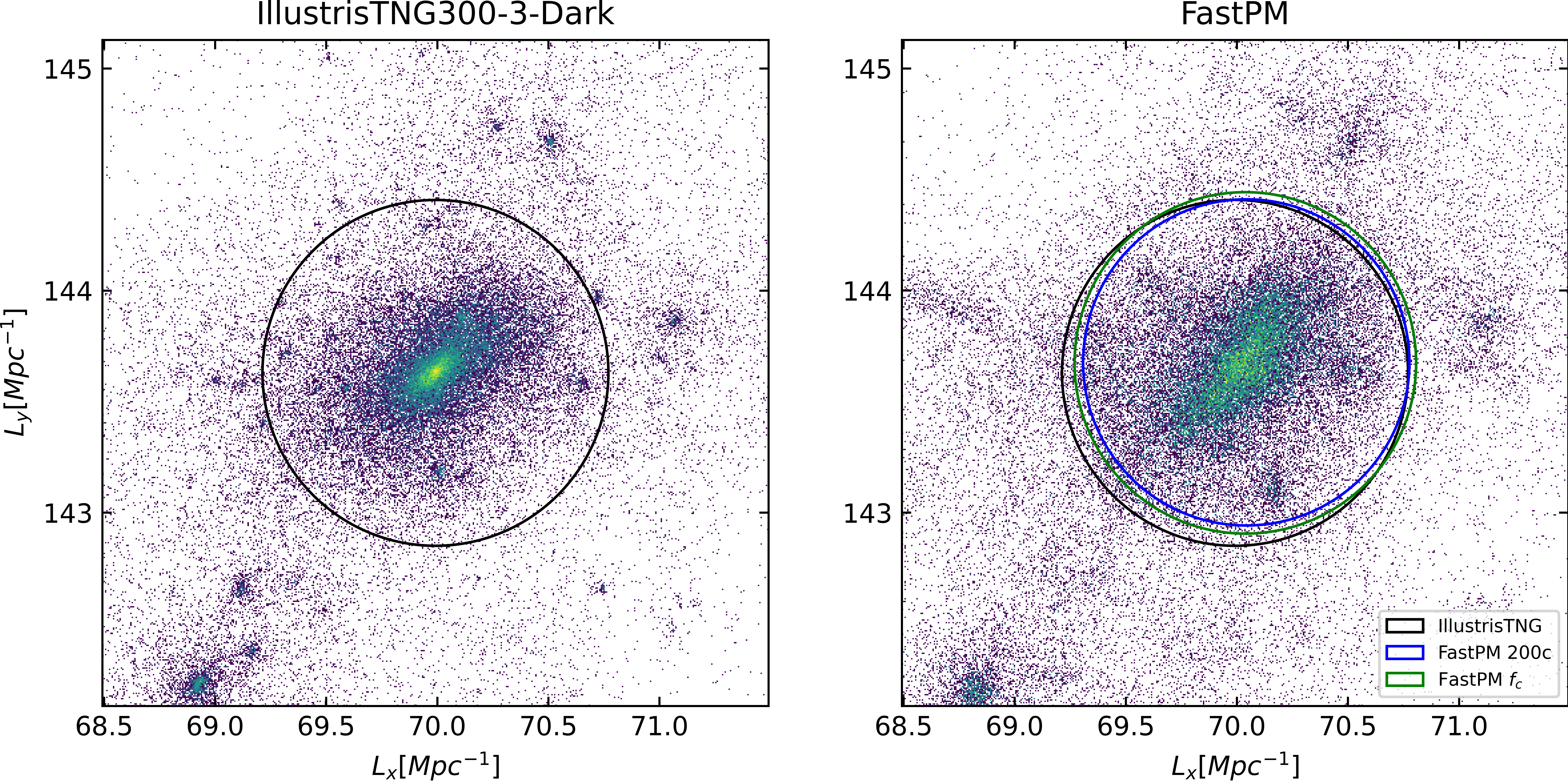}
\caption{Visualisation of the particle distribution in a slice from TNG (left panel) and FastPM (right panel), at the same position in the simulation box. In the left panel, the black circle marks a halo with an enclosed overdensity of 200 times the critical density in TNG. In the right panel, the black circle is the exact same circle as in the left panel, whereas the blue circle gives the boundary of the halo in FastPM when constructed with an equal overdensity value of 200. The green circle, on the other hand, is the corresponding boundary of a halo with overdensity $f_c$, which (on average) contains the same mass as TNG haloes of this size. Unlike for the spherical overdensity mass, the locations and shapes of haloes in FastPM fit well with those in TNG. \label{fig:slice_location}}
\end{figure*}

We opt for a linear time step in the scale factor for our FastPM simulations, which produces better halo statistics than logarithmic time steps \citep{Feng2016}. We ran all FastPM simulations from $z=9$ to $z=0$ in 40 steps. The Fourier mesh size is set to four times the number of particles per dimension,i.e. Nmesh=4. For example, a simulation of $1024^3$ particles has a mesh size of $4096^3$. We have tested cases with different Nmesh, and the results indicate that for Nmesh=4 a best compromise between the obtained halo properties and the invested computational resources is reached.

\begin{table}
	\centering
	\caption{FastPM simulations analog to Illustris-TNG}
	\label{table:simu}
	\begin{tabular}{lcr} 
		\hline
		Particles & $d_{\rm s}/\mpchi$ & Corresponding TNG\\
		\hline
            $480^3$ & 0.427 & TNG300-3-Dark ($625^3$) \\
            $640^3$ & 0.320 & TNG300-3-Dark ($625^3$) \\
            $960^3$ & 0.214 & TNG300-2-Dark ($1250^3$) \\
            $1280^3$ & 0.160 & TNG300-2-Dark ($1250^3$) \\
		\hline
	\end{tabular}
\end{table}

\subsubsection{N-body simulations}

This paper uses the N-body simulations of the IllustrisTNG project (hereafter TNG) \citep{Weinberger2017, Springel2018, Pillepich2018, Nelson2018, Marinacci2018, Naiman2018} as a reference for high-resolution calculations. We are only considering the dark-matter simulations of TNG, which are conducted with the moving-mesh code {\small AREPO} in three different box sizes and with varying resolutions. The TNG simulations were carried out with Planck15 cosmology \citep{PlanckCollaboration2016}, which includes $\Omega_m=0.3089$, $\Omega_b=0.0486$, $h=0.6774$, $n_s=0.9667$, and $\sigma_8=0.8159$. The simulations evolved from redshift $z=127$ to the present day. We focus on the dark-matter-only suite of TNG300-2-Dark and TNG300-3-Dark, which have $1250^3$ and $625^3$ dark matter particles, respectively, in a $205\,h^{-1}{\rm Mpc}$ periodic box. We will refer to these simulations as TNG.

We compare the halo properties between FastPM and N-body simulations by running FastPM simulations with the same initial conditions as TNG. Unfortunately, FastPM cannot work with an uneven number of particles, such as $625^3$. To compensate for this, we run FastPM with slightly different total particle numbers of $480^3$ and $640^3$ for TNG300-3-Dark ($625^3$), and with $960^3$ and $1280^3$ for TNG300-2-Dark ($1250^3$), as indicated in Table~\ref{table:simu}.

For fair comparisons, we applied the phase space halo finder of {\small ROCKSTAR} \citep{Behroozi2013} to FastPM and TNG. We focus on the halo mass definition of $M_{\rm 200c}$, i.e., the average mass density enclosed in a halo is 200 times the critical density, $\rho_{\rm cr}$.

\subsubsection{One-to-one halo matches}
When the same initial conditions are applied, the large-scale structures of the TNG and FastPM simulations agree with each other \citep{Feng2016}. However, the one-to-one halo comparisons between FastPM and full N-body simulations may differ depending on the cosmological parameters and mass resolutions. To quantify the discrepancies on a one-to-one halo basis, we need to identify pairs of cross-matched haloes in the TNG and FastPM simulations, which were both run with the same initial conditions. For a given main halo in TNG with a radius of $R_{\rm 200c}$, we search for all the neighbouring haloes within $2R_{\rm 200c}$ in the FastPM simulation box. If more than one halo is found, we select the most massive halo as the appropriate match.

\begin{figure}
\includegraphics[width=\columnwidth]{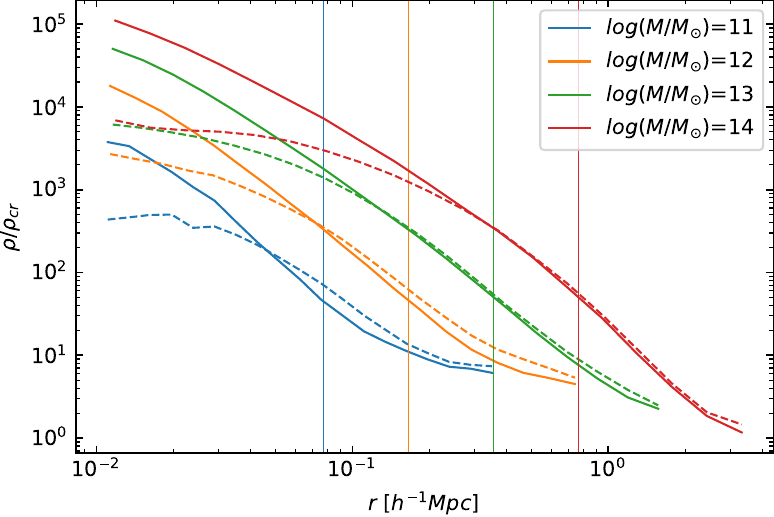}
\caption{Exemplary halo density profiles for TNG (solid lines)  and FastPM (dotted lines) haloes. Colors correspond to different halo mass bins, and the vertical lines are corresponding halo radii $R_{200c}$. The density near the centers of FastPM haloes is washed out and much lower than for the full N-body TNG haloes. The mass missing in the core is instead spread out more smoothly around the edge and outside of the haloes.\label{fig:nfw}}
\end{figure}

In Figure~\ref{fig:slice_location}, we show a halo with a mass of approximately $M_{\rm 200c}\sim10^{14}\msun$ to demonstrate the similarity between FastPM and TNG. The spatial positions and shapes of the haloes in both simulations are comparable. However, the halo masses in FastPM are usually much lower than those in TNG, and the percentage difference in mass decreases with the halo mass. Figure~\ref{fig:nfw} shows examples of halo density profiles for TNG haloes (solid lines) and FastPM haloes (dotted lines) with masses ranging from $M_{\rm 200c}=10^{11}\msun$ to $10^{14}\msun$. It is evident that the halo density profiles in FastPM are significantly flattened towards the halo centres, particularly within $r<0.5R_{\rm 200c}$. This is because, under the same halo definition of $M_{\rm 200c}$, haloes in FastPM have smaller $R_{\rm 200c}$ to reach the density threshold of $200\rho_{\rm cr}$.

\begin{figure}
\includegraphics[width=\columnwidth]{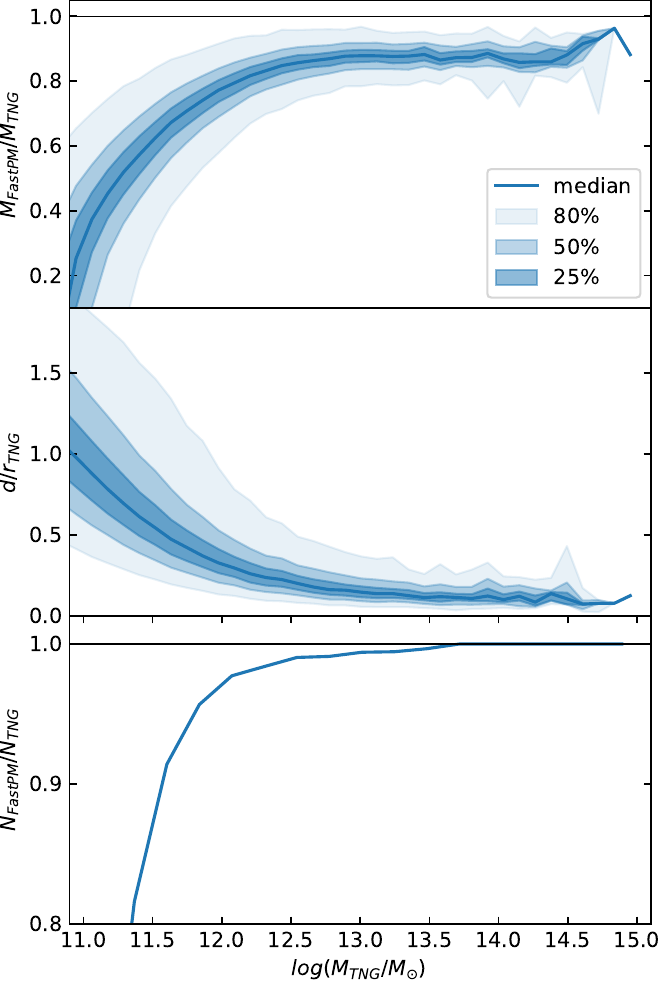}
\caption{{\it Upper panel:} Ratio of the FastPM halo mass and the corresponding TNG halo mass in matched pairs of haloes, as a function of TNG halo mass. The masses of most haloes in the FastPM simulation are measured to be lower than the corresponding ones in the TNG. 
The shaded regions correspond to the 25, 50 and 80 percentiles of the distribution around the median values.} {\it Middle panel:} Distance of the halo centres of pairs of matched haloes in FastPM and TNG simulations, in units of the radius of the halo in the TNG simulation. The blue line marks a running median value. Most haloes in the FastPM simulation are located within the virial sphere of the corresponding TNG halo, and the more massive the halo, the smaller the spatial offsets tend to be. {\it Lower panel:} Fraction of matched halo pairs in the FastPM and TNG simulations that can be considered robust identifications. Even for low particle numbers per halo, more than $\sim90$ percent of the haloes can be robustly matched across the simulations\label{fig:121}
\end{figure}

The upper panel of Figure~\ref{fig:121} displays the ratio of the halo masses of the corresponding pairs when the haloes in TNG are matched with FastPM. It is evident that the masses of most haloes in the FastPM simulation are lower than the corresponding ones in TNG, and the proportion of less massive haloes is higher than that of more massive ones. The middle panel shows the distance of the matched FastPM haloes to the corresponding TNG halo divided by the radius of the TNG halo, with the blue line showing the running median value. For massive haloes, the one-to-one matched haloes in the FastPM simulation are always situated within the virial sphere of the corresponding TNG halo, and the pair distance is always very small. For low-mass haloes, the distances become significantly larger in terms of the TNG halo radius, and there are more haloes located outside of the original TNG halo sphere. Although the average distance remains within the virial sphere of the primary TNG halo, the number of misidentifications probably starts to increase.

To obtain reliable statistics, we only consider halo pairs as valid matches when they are separated by no more than one halo radius and the mass ratio of $M_{\rm TNG}/M$ is between $0.5$ and $2$. The bottom panel of Figure~\ref{fig:121} shows the percentage of one-to-one halo matches that meet this reliability criterion. This fraction decreases with decreasing halo mass. However, for a wide range of halo masses, the percentage of successfully identified one-to-one halo matches is greater than 90\%. As expected, the higher the halo mass, the higher this fraction is.

\subsection{Empirical halo mass correction}
We attempt to match the mass of the haloes in FastPM to those in TNG by drawing spheres from the halo centre in FastPM and calculating the enclosed mass. When the mass inside the sphere is equal to the halo mass in TNG, we assume that the new haloes contain a similar amount of material as in TNG. However, even when the halo masses are equal, other halo properties may still differ.

\begin{figure}
\includegraphics[width=\columnwidth]{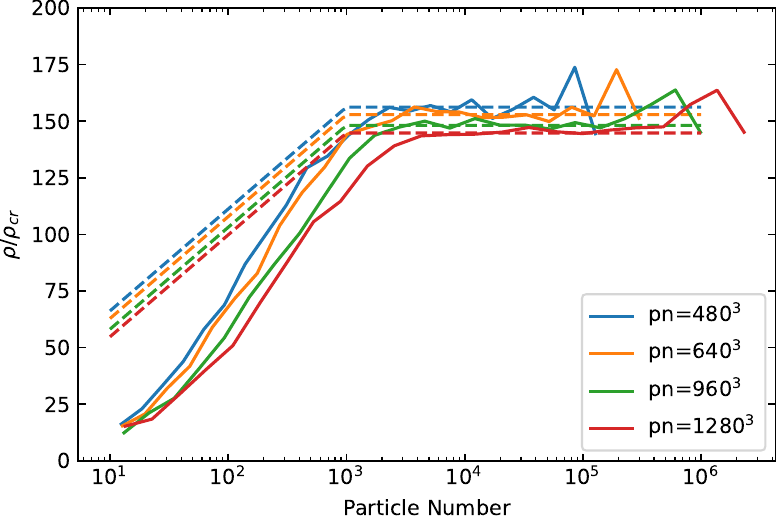}
\caption{The relation between particle number and the effective halo overdensity required to reach in FastPM  the exact same $M_{200c}$ halo mass as found for the corresponding halo in a full N-body simulation. The solid lines are the average values of these overdensities for the mass-matched haloes. The dashed lines give our corresponding fitting formula, which is meant to match this empirical finding, adjusted however at the low particle number end in order to avoid an overcorrection of the halo abundance there.  The different colors correspond to different particle resolutions employed in the FastPM simulations. \label{fig:cc-particle}}
\end{figure}

To gain a more concise representation of the mean correction, we calculated the density inside the newly discovered haloes and studied the connection between the mean enclosed density and the original particle number of the halo. Figure~\ref{fig:cc-particle} shows the relationship between the number of particles and the overdensity of the corrected haloes determined in our FastPM simulation set, with the colours corresponding to the results for different total numbers of particles used for the FastPM simulations, as labelled. We can observe that the inferred density of small haloes is usually lower than that of larger ones, until the haloes become large enough and a plateau appears. We can describe the behaviour of this empirical relation with three parameters: a slope $k$ for the increase part at low masses, a turning point in the number of particles $p$ when an approximately constant value is attained, and a height $h$ for the overdensity value of the emerging plateau. When the number of particles inside the original FastPM halo is $n$, a fitting formula for the average overdensity of the empirically corrected FastPM haloes can be expressed as
\begin{equation}
f_c=
\begin{cases}
k(\log n- \log p)+h& \text{ $ n< p $}\\
h& \text{ $ n\ge p $}
\end{cases}
\label{fitting1}
\end{equation}
where we denote $f_c$ as the overdensity of the corrected halo. For each halo with a particle number $n$ in the original FastPM simulation, we assign an overdensity $f_c$ according to Eq.~(\ref{fitting1}). The new halo mass and its other properties are then calculated from the sphere with an enclosed density of $f_c\rho_{\rm cr}$.

\begin{figure*}
\includegraphics[width=0.7\textwidth]{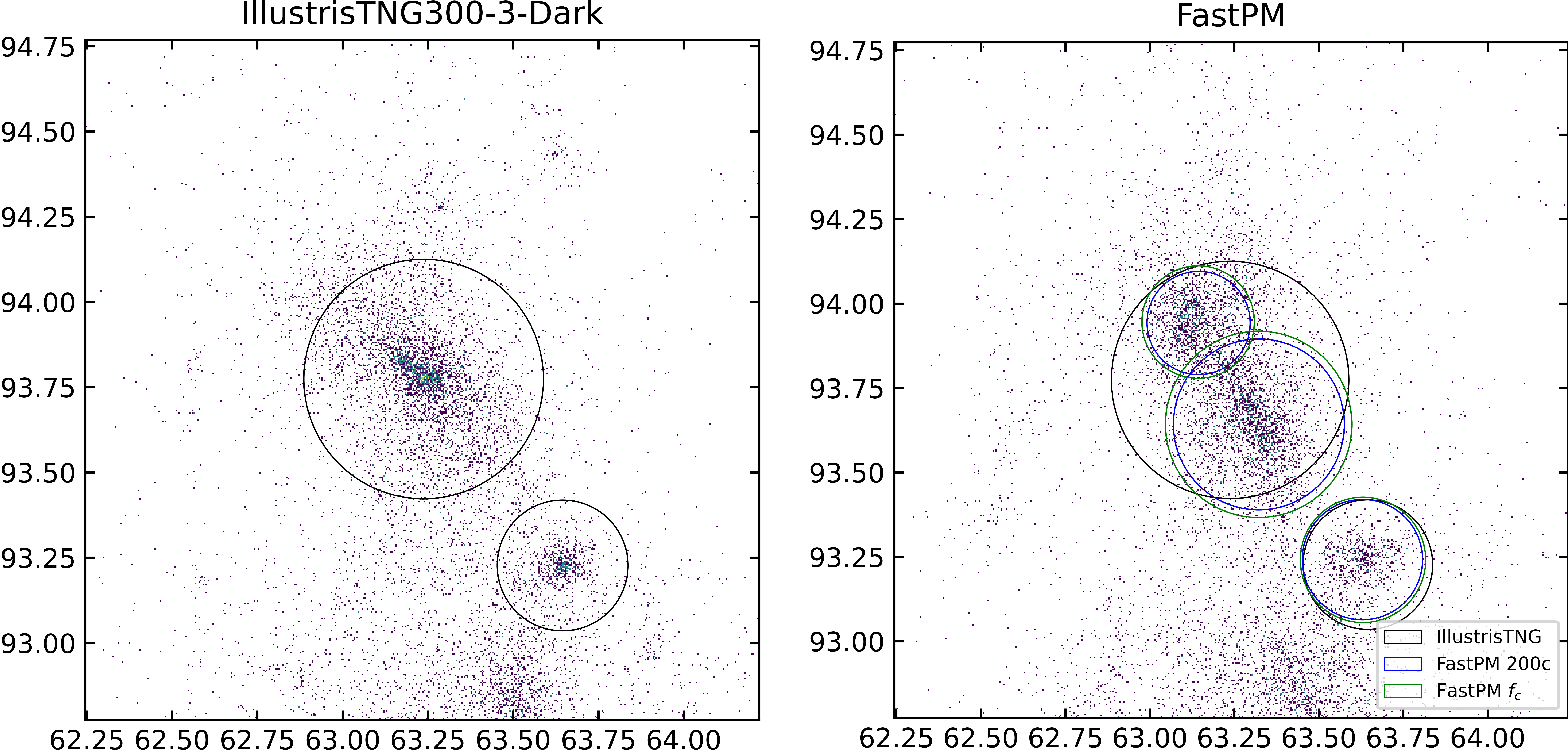}
\caption{Illustration of the spurious haloes in FastPM caused by its lack of small-scale modes. The halo in the centre of the TNG simulation (left panel) is wrongly identified as two smaller haloes in the FastPM simulation (right panel). This is caused by the decreased halo concentration in FastPM as indicated in Figure~\ref{fig:nfw}.} \label{fig:spurious}
\end{figure*}

The parametric form of Eq.~(\ref{fitting1}) can be used to fit the data of the filtered set of halo pairs. However, when applied to all haloes in a FastPM simulation, the resulting mass function does not match that from a full N-body simulation at the low-mass end. This is due to the coarser resolution of the structures in the FastPM box, which leads to a higher rate of spurious haloes and a larger number of smaller haloes that have not yet merged. This effect is clearly illustrated in Figure~\ref{fig:spurious}, where the halo located in the centre of the TNG panel is incorrectly identified as two smaller haloes in FastPM.  To address this problem, we found that changing the slope value $k$ to $k\simeq 45$ and keeping $p=10^3$ (as suggested in Figure~\ref{fig:cc-particle}) produces the best reproduction of the halo mass function and clustering of the TNG haloes.

\begin{figure}
\includegraphics[width=\columnwidth]{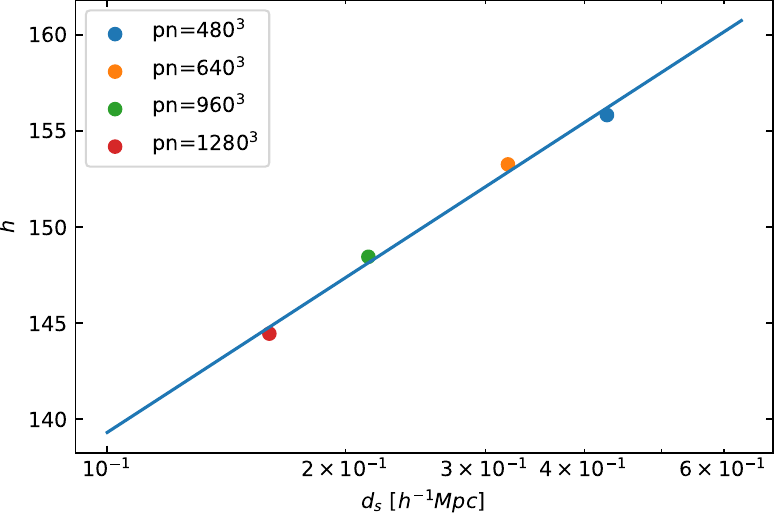}
\caption{Mean values of the overdensity plateau $h$ in Figure~\ref{fig:cc-particle}, for different particle densities of the underlying FastPM simulations. The blue line is a log-linear fit to the measurements, allowing to account for resolution dependence in Eq.~(\ref{fitting1}). \label{fig:height}}
\end{figure}

The height of the plateau $h$ is slightly influenced by the resolution of the FastPM simulation, as demonstrated in Figure~\ref{fig:height}. This figure shows the connection between $h$ and the average particle distance of the simulation $d_{\rm s}$, with the following fitting formula, 
\begin{equation} 
h=26.81\log(d_{s}/\mpchi) + 166.11. \label{h-ds} 
\end{equation} 
The resolution of the simulation is represented by $d_{\rm s}$, and it appears that the height $h$ is directly proportional to the logarithm of the mean particle separation of the FastPM simulation. This implies that a fitting formula for $f_c$ can be easily derived when the particle number density is changed. It is important to note that the configuration of the FastPM parameters has a significant impact on the fitting parameters. We also emphasise that the fitting parameters depend on the adopted Nmesh value, which we set as four times the particle number per dimension. In this sense, the force resolution and the mass resolution are coupled in this study.

\begin{figure}
  \includegraphics[width=\columnwidth]{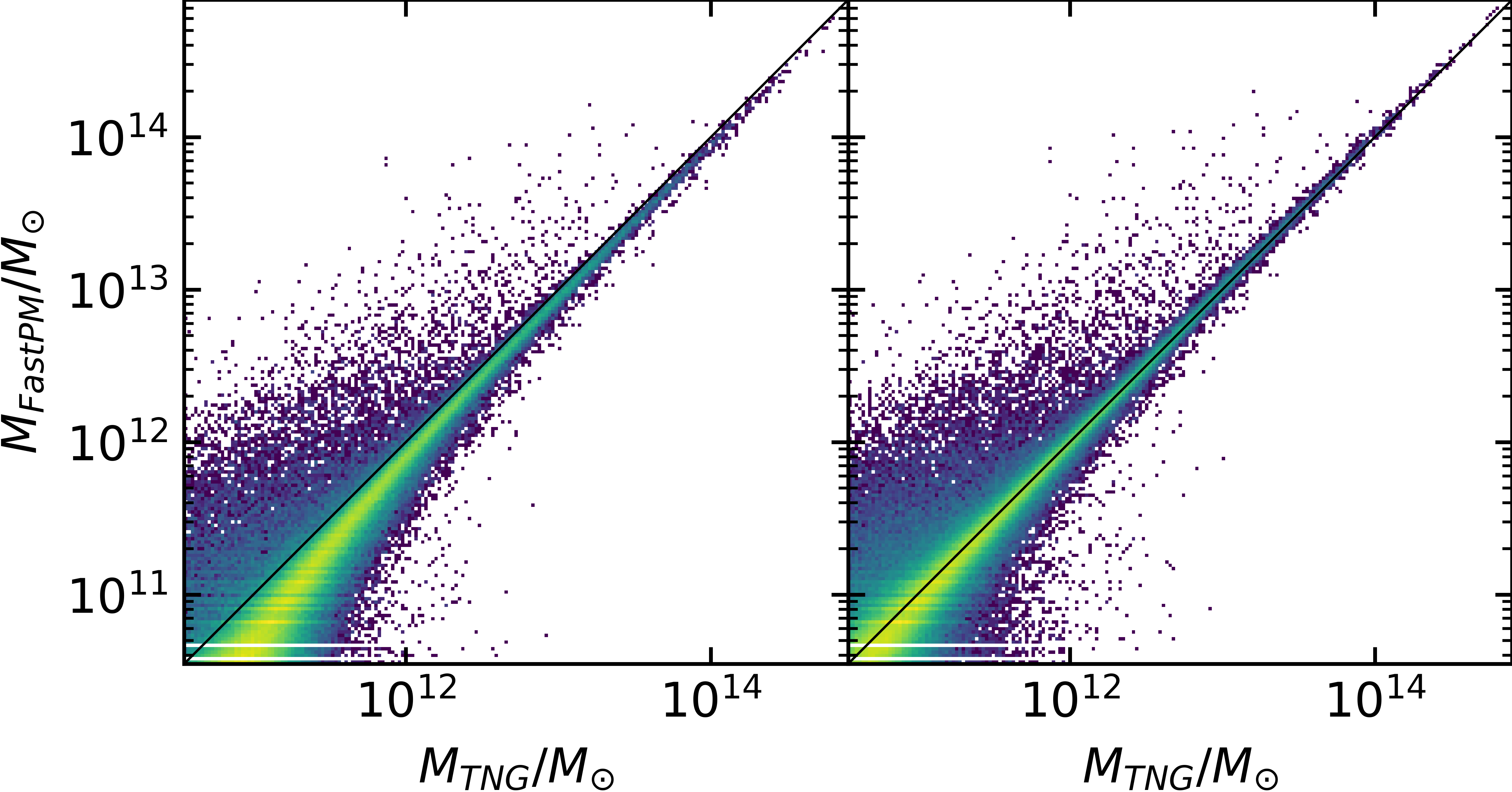} \caption{{\it Left panel:} Comparison of the spherical overdensity halo masses obtained for FastPM and TNG simulations in matched halo pairs when using the canonical overdensity of $200$ with respect to critical in both cases. {\it Right panel:} Here the FastPM masses with overdensity $f_c$ are compared to the TNG halo masses. The oblique black line marks the identity relation in both cases. The $f_c$-corrected halo masses fit the TNG halo masses rather well, as desired. \label{fig:mass}}
\end{figure}

\subsection{$f_c$-corrected halo catalogues}

We can use the {\small ROCKSTAR} code \citep{Behroozi2013} to find the particle groups and the spherical overdensity mass of the haloes, with the empirically determined halo overdensity $f_c$. This public code, which usually employs a constant overdensity value, can be easily modified to determine a variable spherical overdensity mass $f_c$ instead by changing its {\tt properties.c} file. Our modification replaces the original definition of the halo mass $M_{500c}$ in {\small ROCKSTAR} with our $f_c$ corrected value, while the original standard $M_{200c}$ is still kept for comparison.

We compare the standard $M_{200c}$ halo catalogues of FastPM and N-body simulations to validate the $M_{f_{c}}$ halo catalogues constructed for FastPM. Taking into account the limited force resolution and coarse timestepping of the approximate methods, we only analyse the properties of individual haloes without considering subhaloes. Figure~\ref{fig:mass} shows a comparison of pairs of matched haloes, considering the FastPM masses of the $M_{200c}$ halo (left panel) and the $M_{f_c}$ halo (right panel) compared to the spherical overdensity masses of the corresponding haloes in TNG. The line with unit slope marks the identity. It is evident that the $M_{f_c}$ halo masses fit the TNG haloes quite well from the low-mass end to the high-mass end, while the plain $M_{200c}$ halo masses are always slightly lower than those found in the TNG simulation. The matching works quite well, especially at the high-mass end, while for low-mass haloes, a larger number of spurious haloes and haloes that are not identified in the FastPM simulation create substantial scatter.

\begin{figure}
\includegraphics[width=\columnwidth]{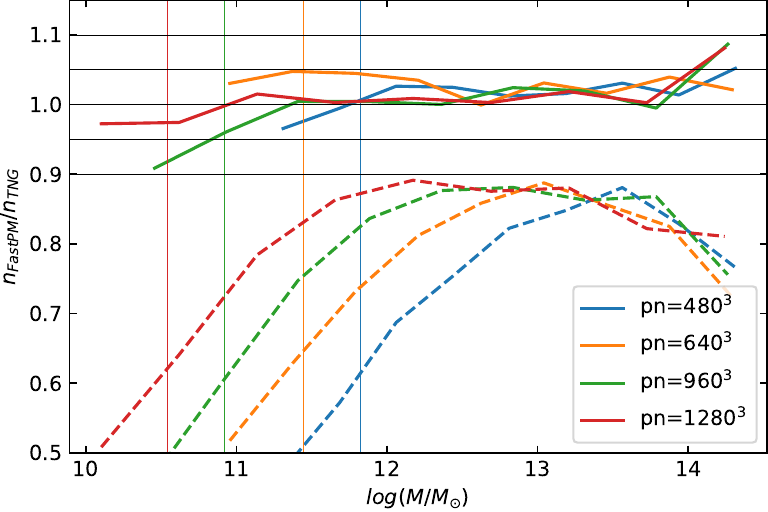}
\caption{Halo mass function compared between FastPM and TNG halo catalogues for the same initial conditions. The dashed lines are for halo masses determined with canonical spherical overdensities of 200 with respect to the critical density. In contrast, the solid lines use $f_c$ corrected halo masses in the case of the FastPM simulations. The vertical lines mark the masses of fiducial 100 particle haloes for the different FastPM simulations, which are labelled in the color key with a number that gives the cube root of the total particle number in the corresponding simulation models. The precision with which the $f_c$-corrected haloes recover the mass function of full N-body simulations lies within 5 percent over nearly the full halo mass range. \label{fig:hmf}}
\end{figure}

A statistical comparison is presented in Figure~\ref{fig:hmf}, which compares the halo mass functions of four FastPM simulations with those of their corresponding TNG simulations. FastPM runs of $480^3$ and $640^3$ particles are compared with TNG-3, while FastPM runs of $960^3$ and $1280^3$ are compared with TNG-2. The dashed lines indicate the results when the ordinary $M_{200c}$ halo masses are used, whereas the solid lines are based on the $f_c$ halo masses for FastPM runs. The thin vertical lines mark halo masses corresponding to 100 particles for the four different FastPM simulations. It is evident that the $M_{200c}$ halo catalogues have a large discrepancy compared to the TNG catalogues, the difference in the halo number density being greater than 10\% even for the most massive haloes. In contrast, the precision of the $M_{f_c}$ catalogues is much improved, and the halo density agrees with TNG to within 5\% for almost the entire mass range, except for slightly larger deviations at the low and high mass ends.

\begin{figure}
\includegraphics[width=\columnwidth]{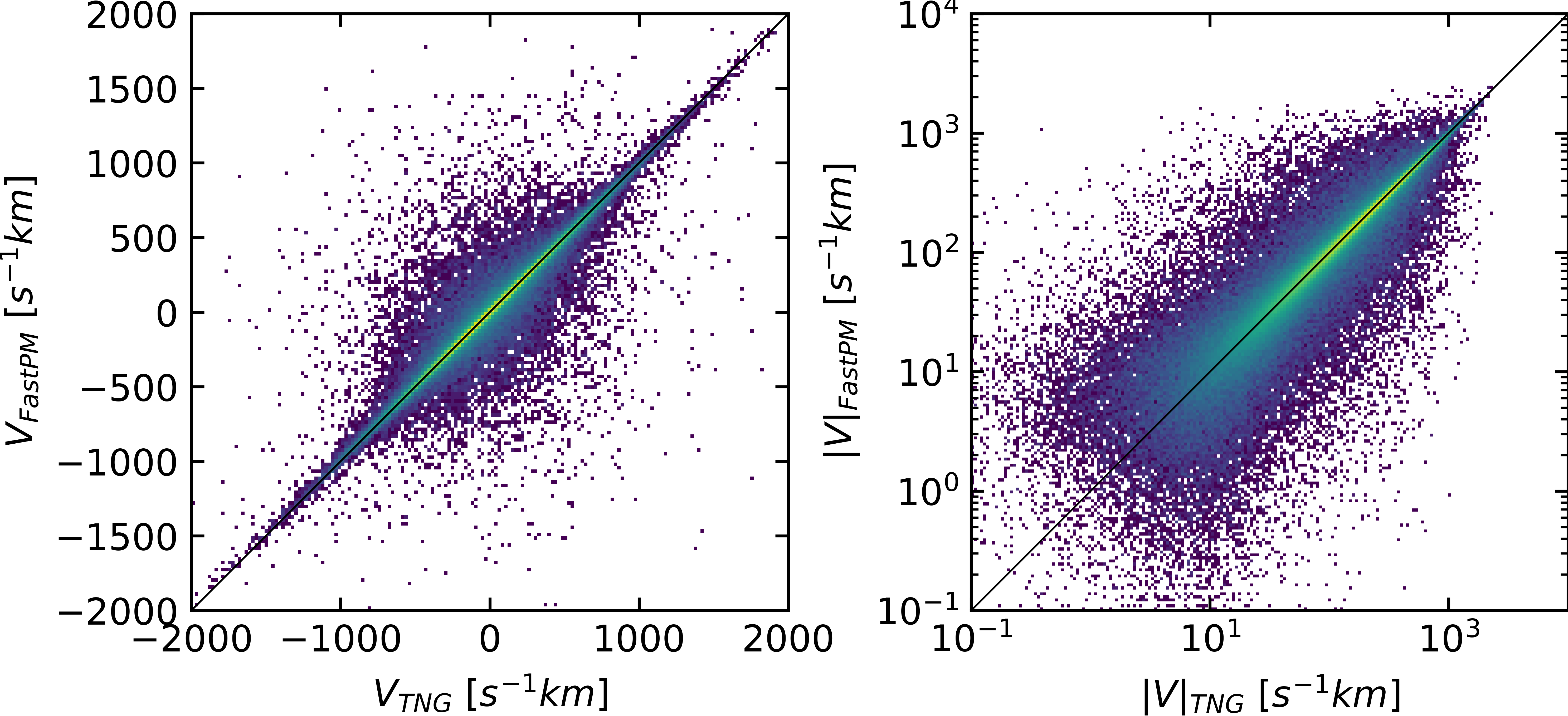}
\caption{Halo bulk velocity comparison for one-to-one matched pairs of haloes in FastPM and full N-body simulations. In the left panel, the velocity along a  randomly chosen direction is shown, whereas  the right panel compares the absolute values of the halo velocities. \label{fig:velocity}}
\end{figure}

\begin{figure}
\includegraphics[width=\columnwidth]{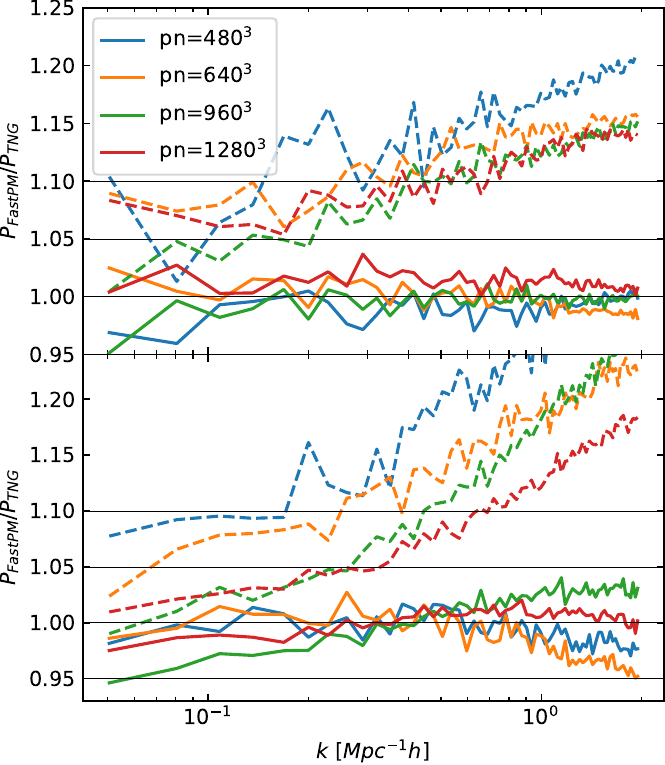}
\caption{Halo clustering power spectrum comparison between FastPM and TNG at redshift $z=0.3$. The line styles are the same as in Fig.~\ref{fig:hmf}. The top panel shows the power spectrum in real space while the bottom panel compares the power spectrum in redshift space. The precision of the results for the $f_c$-based halo catalogues are improved to lie within 5 percent of the full N-body results for both types of power spectra. \label{fig:pk}}
\end{figure}

\begin{figure}
\includegraphics[width=\columnwidth]{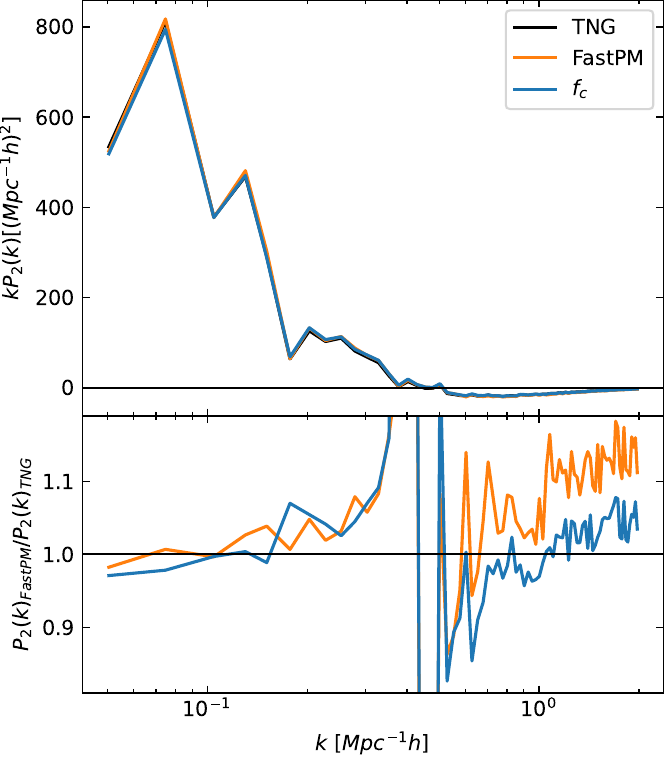}
\caption{Comparisons of the quadrupole moment ($P_2(k)$) between FastPM and TNG at redshift $z=0.3$ in redshift space. The dark, yellow, and blue lines correspond to the outcomes of TNG, FastPM, and $f_c$, respectively. The top panel illustrates the power spectrum quadrupole moments, while the bottom panel displays the ratios of $P_2(k)_{\rm FastPM}/P_2(k)_{\rm TNG}$ for the original FastPM and $f_c$ results. }\label{fig:multpole}
\end{figure}

We also analyse the halo clustering power spectrum in both real- and redshift spaces for the corrected halo catalogues. Figure~\ref{fig:velocity} shows the comparison of the bulk velocities of the haloes in FastPM $1280$ runs and the corresponding haloes in TNG-2. We randomly select one spatial direction vector for each halo and compare the velocity along this direction in the left panel, as well as the absolute value of the full velocity in the right panel. The results show that the value and individual components of the velocity match well with the TNG on average, so the velocity can be considered reliable for placing the haloes into redshift space.

We can compare the halo power spectra of FastPM and full N-body simulations by selecting haloes with a mass of more than 100 particles for each FastPM simulation. The results are shown in Figure~\ref{fig:pk}, with the upper panel showing the power spectrum in real space and the lower panel showing the corresponding measurement in the redshift space at redshift $z=0.3$. The dashed lines represent the results for the selected $M_{200c}$ haloes, while the solid lines represent the selected $M_{f_c}$ haloes. Compared to the original $M_{200c}$ halo catalogues, the power spectrum of the $f_c$ halo catalogues shows a significant improvement and is within the accuracy 5\% for the entire range of wave numbers. Even at the smallest scale, the precision is still better than 2 percent for the power spectrum in both real and redshift spaces.

In the analysis of the halo power spectrum in redshift space, multipole moments are of significant importance. Therefore, we present the results for the quadrupole moment of the redshift-space halo power spectrum ($P_2$) in Figure~\ref{fig:multpole}. Considering the consistent outcomes across different particle numbers of $pn=480$, $640$, $960$, and $1280$, we only display the results for $pn=1280$, for clarity. It can be observed that both the corrected FastPM and the $f_c$ method closely align with the trends exhibited by the TNG model for the quadrupole moment. However, unlike for the real space and redshift space power spectra, the $f_c$ method does not yield significant improvement in the multipole moments on large scales, but it is clearly better than the original FastPM results on small scales of $k>1\,h\,{\rm {Mpc}^{-1}}$. The ratios of $P_2(k)_{\rm FastPM}/P_2(k)_{\rm TNG}$ for the original FastPM and $f_c$ results are shown in the bottom panel as the yellow and blue lines, respectively. The large discrepancies at around $k\sim0.5\,h\,{\rm {Mpc}^{-1}}$ are simply caused by the small values of $P_2(k)_{\rm TNG}$ that is approaching zero at these scales.

\section{Application to different cosmologies}\label{section:cosmology}

In Section~\ref{section:method}, we developed the $f_c$-correction method to substitute the standard $M_{200c}$ halo catalogues in FastPM simulations with modified ones that are more similar to the results of full N-body simulations such as IllustrisTNG-300-Dark. We now want to investigate whether this method can be used in other contexts, particularly for different cosmologies, as this would enable the utilisation of cost-effective FastPM simulations in cosmological inference. We conducted five sets of simulations with varying cosmological parameters, which is a much wider range than what is currently accepted by cosmological constraints. Full N-body simulations of this validation set were run with the {\small AREPO} code, just as the IllustrisTNG series, while FastPM runs used default settings. The five pairs of simulations had the same box size of $205\mpchi$ and a total particle number of $512^3$ to reduce computational cost. The cosmological parameters of the models are listed in Table~\ref{table:cosmology} for reference.

\begin{table}
	\centering
	\caption{Cosmology parameters used in validation runs}
	\label{table:cosmology}
	\begin{tabular}{lccr} 
		\hline
		Simulation & $\Omega_{\rm m}$ & $h$ & $\sigma_8$\\
		\hline
            s1 & 0.24 & 0.66 & 0.70\\
            s2 & 0.26 & 0.68 & 0.75\\
            s3 & 0.28 & 0.70 & 0.80\\
            s4 & 0.30 & 0.72 & 0.85\\
            s5 & 0.32 & 0.74 & 0.90\\
		\hline
	\end{tabular}
\end{table}

\begin{figure}
\includegraphics[width=\columnwidth]{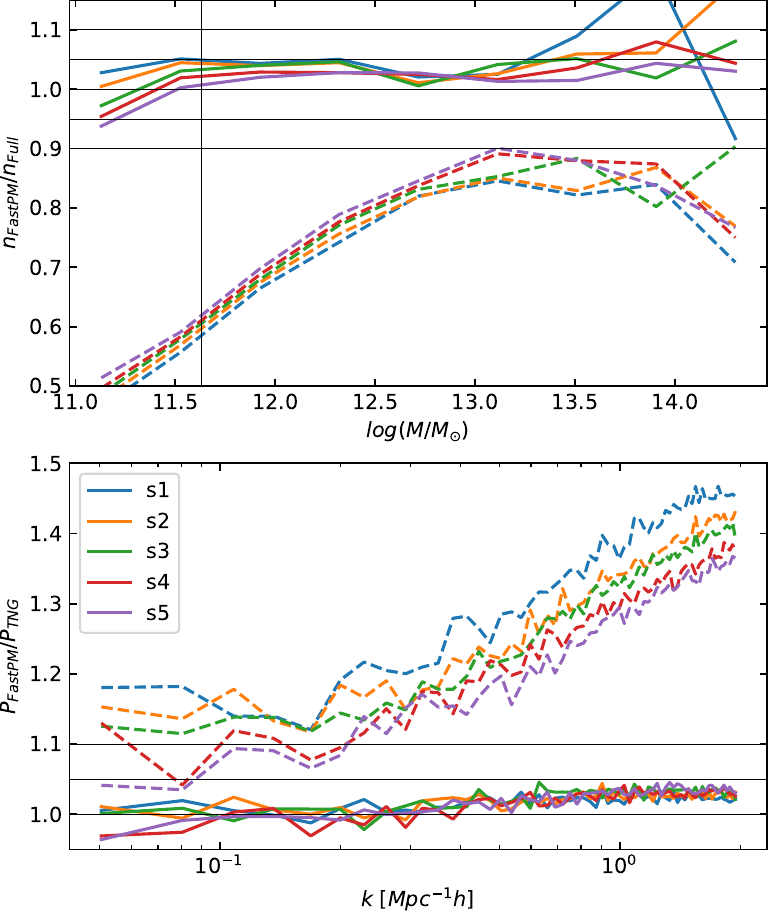} \caption{Comparison of halo mass functions (upper panel) and halo power spectra (lower panel) for 5 sets of simulations with different cosmological parameters, as listed in Table~\ref{table:cosmology}. The dashed lines are the results of ordinary halo catalogues based on an overdensity of 200 relative to the critical density, whereas the solid lines are for $f_c$-overdensity values in the FastPM halo catalogues. For the 5 sets of cosmology settings explored, the precision of the halo mass function and the power spectrum stays (almost) always within 5 percent, implying that the correction method can be applied independently of cosmology for a reasonable range around the $\Lambda$CDM concordance model. \label{fig:cosmology}}
\end{figure}

For each pair of simulations with the same cosmological parameters and initial conditions, we apply the $f_c$-correction method to the FastPM simulations to obtain halo catalogues with plain $M_{200c}$ and corrected $M_{f_c}$ halo masses. Figure~\ref{fig:cosmology} displays the comparison results between the haloes in the full $N$-body runs and the corrected halo catalogues. The upper panel shows the comparison for the halo mass function, while the lower panel shows the result for the halo clustering power spectrum in real space. The vertical black lines in the upper panel represent a fiducial halo mass composed of 100 particles. As the particle mass differs slightly in different cosmologies, we just draw one vertical line for simplicity. The same mass limit has also been applied to select the haloes used for calculating the clustering power spectra shown in the lower panel. The precision achieved for the $f_c$ halo catalogue is consistent with that found in Section~\ref{section:method}, and both the halo mass function and the power spectrum agree within 5\% with the full $N$-body results. For the model `s1', the accuracy of the correction for the halo mass function is slightly worse at the massive end, but this could still be due to small number statistics. Nevertheless, these results demonstrate that the empirically calibrated $f_c$ correction method can be applied to different cosmologies over a wide range of cosmological parameters without requiring a renewed calibration.

\section{Redshift dependence}\label{section:redshift}

Previously, we discussed why approximate methods are much faster than full N-body simulations. This is because they only require tens of steps, whereas full N-body simulations require thousands. In addition, FastPM uses comparatively low force resolution (plain PM instead, e.g., TreePM) which is faster to compute. In the setup studied, FastPM only needs 40 steps, which are linearly spaced from $a=0.1$ to $a=1$. This means that FastPM has very few steps at higher redshifts. In fact, it only takes 17 steps to reach $z=1.0$, suggesting that statistics at high redshifts may not be as accurate as at $z=0$. 

\begin{figure}
\includegraphics[width=\linewidth,scale=1.0]{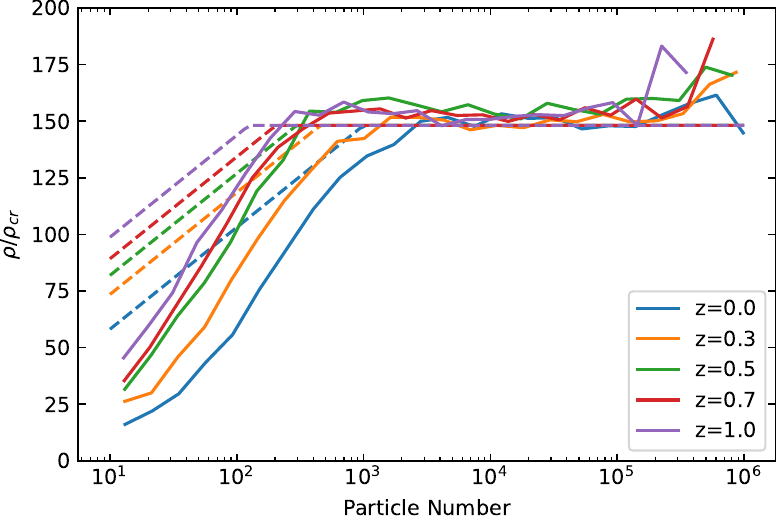}
\caption{Similar to Fig.~\ref{fig:cc-particle}, but for different redshifts, as labelled by the  colors. The solid lines give the average overdensity values needed for FastPM haloes to recover the masses expected based on matched pairs of TNG haloes, while the dashed lines are our corresponding correction formulae, tuned to avoid an overcorrection of the halo abundance at the low mass end. \label{fig:redshift fitting}}
\end{figure}

We analyse five redshifts: $z = 0.0$, $0.3$, $0.5$, $0.7$, and $1.0$, which cover the most important redshift range for cosmological studies with galaxy surveys. Applying the same methods as in Section~\ref{section:method}, Figure~\ref{fig:redshift fitting} shows the relationship between particle number and the mean effective overdensity of the mass-matched haloes identified at different redshifts. The solid lines represent the average results for the matched pairs of haloes and demonstrate three distinct systematic characteristics. Firstly, the slope of the increase at low masses, $k$, is similar for different redshifts. Secondly, the height of the plateau, $h$, remains constant. Third, the turning point moves to a lower particle number with increasing redshift. Therefore, we use the same slope $k$ and height $h$ as for the redshift $z=0$, but we simply multiply the turning point by the cubic scale factor $a^3$, to account for the variation in the redshift of the turning point. The corresponding fitting formula can thus be expressed as
\begin{equation}
f_c=
\begin{cases}
k[\log n - \log(pa^3)]+h& \text{ $ n< pa^3 $}\\
h& \text{ $ n\ge pa^3 $.}
\end{cases}
\label{eq:fitting2}
\end{equation}

\begin{figure}
\includegraphics[width=\columnwidth]{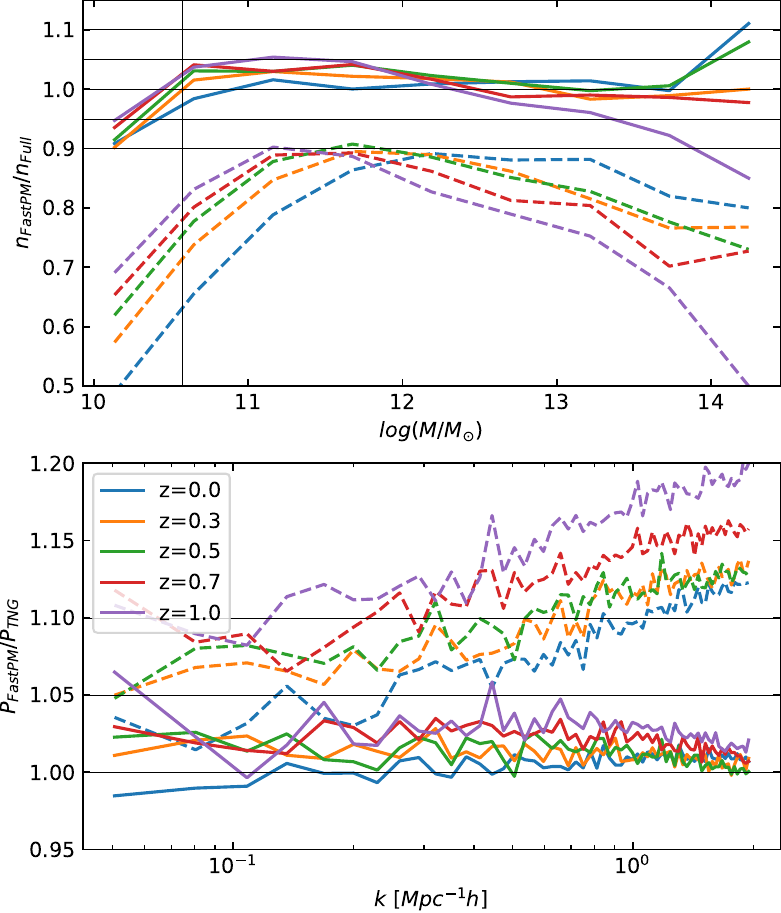}
\caption{Comparison of halo mass functions (upper panel) and halo power spectra (lower panel) between FastPM and full N-body simulations for different redshifts. The dashed lines are the results of halo catalogues with a spherical overdensity of 200, while the solid lines are for $f_c$-corrected FastPM halo catalogues. The precision reached with the $f_c$-based method is about 5 percent or better for the different redshifts, except for the highest mass end in the halo mass function at redshift $z=1.0$. The precision of the halo clustering power spectra  is improved to be within 5 percent for all examined redshifts. \label{fig:redshift}}
\end{figure}

The dashed lines in Figure~\ref{fig:redshift fitting} illustrate the fit results, where a shallower slope was deliberately used to prevent an excessive correction of the halo mass function at the low mass end. The halo mass function and the halo power spectra for the corrected catalogues are shown in Figure~\ref{fig:redshift}. For the original $M_{200c}$ halo catalogues, the halo mass function and power spectra generated by FastPM become worse with increasing redshift. Although the precision of the corrected results also decreases somewhat at higher redshifts, we still find that they meet the accuracy goal of 5\% for both the halo mass function and the halo power spectrum. Compared to results at other redshifts, the results for $z=1.0$ are less accurate, particularly for the massive halo end in the halo mass function but also for the clustering power spectrum. However, this is not unexpected, since FastPM only invests 17 steps to obtain the results at redshift $z=1.0$, and thus it is expected that the halo mass function has a poor accuracy for most non-linear objects.

\section{Halo Occupation Distribution Modeling}\label{section:HOD}

The halo occupation distribution (HOD) approach is a statistical technique used to model the distribution of galaxies within dark matter haloes \citep[see e.g.,][]{Jing1998, Berlind2002, Cooray2002, Yang2004, Zheng2007, Zehavi2011, Guo2015}. The method prescribes the probability of a dark matter halo of a given mass hosting a certain number of galaxies. HOD models have been used to investigate a variety of astrophysical phenomena, from understanding galaxy clustering and the shape of galaxy power spectra to tracing the evolution of cosmic structures over time. Additionally, HOD models are an essential tool for interpreting the clustering patterns of galaxies in large-scale surveys, giving us insight into the underlying cosmological framework and the relationship between dark matter and luminous structures in the universe.

In order to model the clustering of galaxies, we use the HOD approach and the parameterisation proposed by \citet{Zheng2007}. This involves splitting the mean occupation function, $\langle N_{\rm tot}(M)\rangle$, which is the average number of galaxies in a given sample located in haloes of mass $M$, into two components, central galaxies $\langle N_{\rm cen}(M)\rangle$ and satellite $\langle N_{\rm sat}(M)\rangle$ galaxies, which are expressed as follows,
\begin{align} 
\langle N_{\rm tot}(M)\rangle&=\langle N_{\rm cen}(M)\rangle+\langle N_{\rm sat}(M)\rangle,\\ 
\langle N_{\rm cen}(M)\rangle&=\frac{1}{2}\left[1+{\rm erf} \left(\frac{\log M- \log M_{\rm min}}{\sigma_{\log M}}\right)\right],\\ 
\langle N_{\rm sat}(M)\rangle&=\langle N_{\rm cen}(M)\left(\frac{M-M_0}{M^\prime_{1}}\right)^\alpha . 
\label{hod equation} 
\end{align} 

This model has three parameters for satellite galaxies: the cutoff mass scale $M_0$, the normalisation mass scale $M^\prime_{1}$, and the power-law slope $\alpha$ at the high-mass end. The central galaxy occupation function is characterised by the cutoff halo mass $M_{\rm min}$, and the scatter between the galaxy luminosity and the halo mass $\sigma_{\log M}$. For testing purposes, we used the values of these HOD parameters for galaxy samples with $r$-band luminosity thresholds $M_r<-19$, $-20$, and $-21$ listed in Table~2 of \cite{Guo2015}. The halo masses are all expressed in units of $\msunh$. We then employ the Python package {\small Halotools} \citep{Hearin2017} to generate our galaxy catalogues and measure the projected galaxy correlation function $w_p$.

\begin{figure}
\includegraphics[width=\columnwidth]{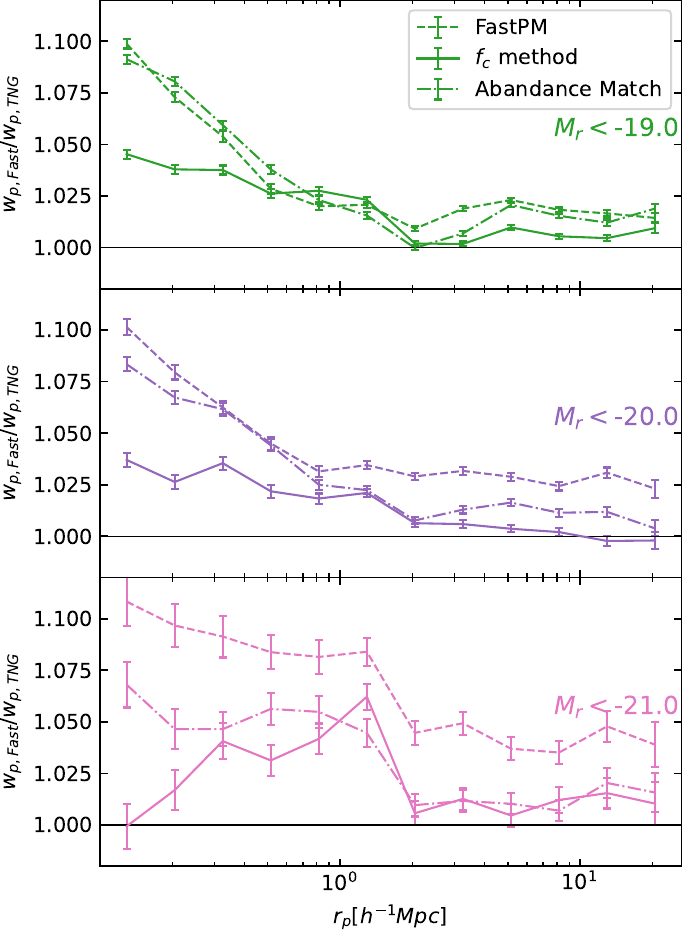}
\caption{
The ratios of $w_{\rm p,Fast}/w_{\rm p,TNG}$ for the original FastPM (dashed lines), the $f_c$ method (solid lines) and the abundance matching method (dot-dashed lines). The error bars denote standard errors from counting statistics. From the top to bottom, the three panels represent the results for the three luminosity threshold samples of $M_r<-19$, $M_r<-20$ and $M_r<-21$, respectively. We can observe that the abundance matching method shows some improvement over the original halo catalogue, particularly in enhancing the representation of high-luminosity galaxies. However, it does not significantly enhance the statistical properties of low-luminosity galaxies. When the $f_c$ method is employed, the accuracy of the measurement is improved on all scales. }\label{fig:hod}
\end{figure}

To further facilitate comparisons, we also applied the abundance matching method. We first rank order all the FastPM haloes by their uncorrected masses and then use the abundance matching method with the TNG simulation to assign the ``corrected'' halo masses. We compare the HOD model predictions from the original FastPM, the $f_c$ method, and the abundance matching results in Figure~\ref{fig:hod}. For each set of HOD parameters, we generate 100 mock galaxy catalogues with different random number seeds to estimate the measurement errors for the TNG300-2-Dark simulation, the original FastPM halo catalogue, the abundance matching halo catalogue and the $f_c$ corrected halo catalogue with $1280^3$ particles (see Table~\ref{table:simu}). The mean galaxy number density for each of the three luminosity thresholds in each simulation is presented in Table~\ref{table:hod}. The galaxy number density in the original FastPM halo catalogues is usually more than 10\% lower than in the N-body simulation, while the $f_c$-corrected method improves the accuracy to within 1 percent or better. As for the abundance matching method, the number density almost equals the one in the TNG simulation.

\begin{table}
	\centering
	\caption{Mean galaxy number density (in units of $10^{-3}h^3\,{\rm Mpc}^{-3}$) for different samples.}
	\label{table:hod}
	\begin{tabular}{lcccc} 
		\hline
		$M_r$ & TNG  & FastPM & $f_c$ & Abandance Match\\
		\hline
            $<-19.0$ & $ 11.15 \pm 0.02 $ & $ 9.81 \pm 0.01 $ & $ 11.18 \pm 0.02 $ & $ 11.15 \pm 0.02 $\\
            $<-20.0$ & $ 5.08 \pm 0.01 $ & $ 4.49 \pm 0.01 $ & $ 5.14 \pm 0.01 $ & $ 5.08 \pm 0.01 $\\
            $<-21.0$ & $ 0.99 \pm 0.01 $ & $ 0.86 \pm 0.01 $ & $ 1.00 \pm 0.01 $ & $ 0.99 \pm 0.01 $\\
		\hline

        \end{tabular}

\end{table}

We computed the projected galaxy correlation function $w_{\rm p}(r_{\rm p})$ by integrating the 3D two-point correlation function up to a maximum distance of $\pi_{\rm max}=40\mpchi$ as described in \cite{Guo2015}. We show the ratios of $w_{\rm p,Fast}/w_{\rm p,TNG}$ for the original FastPM (dashed lines), the $f_c$ method (solid lines) and the abundance matching method (dot-dashed lines) in Figure~\ref{fig:hod}. The error bars denote standard errors from counting statistics. The three panels are assigned to the three luminosity thresholds of $M_r<-19$, $M_r<-20$ and $M_r<-21$, respectively. We can observe that the abundance matching method shows some improvement over the original halo catalogue, particularly in enhancing the representation of high-luminosity galaxies. However, it does not significantly enhance the statistical properties of lower-luminosity galaxies. When the $f_c$ method is employed, measurements on both large and small scales are significantly improved. The deviation from the TNG simulation is around $1\%$ on large scales and less than 5\% on small scales. This finding confirms that it is possible to use the empirically calibrated $f_c$-correction approach for HOD schemes. This thus demonstrates that this method is an effective tool for creating mock galaxy catalogues quickly.

\section{Discussion}\label{section:discussion}

In principle, the $f_c$ correction will have an impact on more than just the halo mass, such as the halo radius, the scale radius, the spin parameter and the formation time. We discovered that the halo radius $R_{200c}$ associated with $M_{200c}$ was smaller in the original simulation than in the full N-body simulations. However, with the $f_c$ correction, the new halo radius was more consistent with the full $N$-body value. Although reasonable estimates of the halo radius were obtained, the halo scale radius and concentration parameter in the FastPM simulations cannot be accurately determined as a result of insufficient small-scale resolution.

We also compare the spin parameter ($\lambda$) in the original and corrected FastPM halo catalogues and find that the agreement with the full $N$-body results is improved with the $f_c$ correction. However, for the halo formation time parameter (defined as the epoch when the halos reach half of their peak mass values in the whole merger trees), both the corrected and uncorrected FastPM halos show considerable scatters compared to the full $N$-body measurements. This implies that the halo formation time should be used with caution in the FastPM simulations.

This present work only focused on the FastPM technique, however, simulations created with L-PICOLA \citep{Howlett2015} should have similar features and should be able to be adjusted with a comparable correction approach. We also expect that halo catalogues produced with COLA could be corrected with the direct method presented here, modulo a possible need for a recalibration. Since the subhalos are not well resolved in such rapid simulation techniques, including the substructures will necessitate extra efforts to calibrate the halo catalogues. We postpone this exploration to future work.

\section{Conclusions}\label{section:Summary}

This paper proposes a simple solution to rectifying the mass definition of haloes identified in approximate simulations, such as FastPM. Our goal is to make the halo catalogues generated from these cost-effective simulations more similar to those from full N-body simulations, particularly in terms of the halo mass function and halo clustering. If successful, the halo catalogues can be used as input for galaxy assignment models, such as HOD modelling, to create more accurate mock galaxy surveys. Our conclusions can be summarised as follows.

(i) The halo masses in the FastPM simulations are significantly underestimated at the low-mass end. Without any corrections, the halo mass functions will be underestimated by more than 10\% for low mass haloes (Figure~\ref{fig:121}). 

(ii) We propose a correction method for the definition of halo overdensity (Equation~\ref{eq:fitting2}), which is based on comparing the halo masses from the FastPM and full $N$-body simulations of IllustrisTNG suites. This method is accurate (within 5\% for most cases) and can be used to obtain accurate mass definitions for the FastPM simulation outputs at different redshifts (up to $z=1$) and for different cosmologies.

(iii) We tested simulations with different resolutions and found that the halo power spectra measured at various redshifts and cosmologies using the $f_c$-corrected FastPM haloes are in agreement with the full $N$-body results, as demonstrated in Figures~\ref{fig:cosmology} and \ref{fig:redshift}.

(iv) We further tested the HOD modelling approach on the FastPM haloes that had been adjusted with the $f_c$-correction. We discovered that, with the $f_c$-correction, the FastPM haloes could reproduce the clustering results of the full $N$-body simulations with an accuracy of up to $5\%$. This is a considerable improvement over the original FastPM halo catalogues in terms of small-scale clustering measurements. This shows that the correction method proposed in this work can be very beneficial in creating galaxy mock catalogues for various cosmologies and redshifts for upcoming surveys.

\section*{Acknowledgements}
  
This work is supported by the National SKA Program of China (grant No. 2020SKA0110100), the CAS Project for Young Scientists in Basic Research (No. YSBR-092) and the science research grants from the China Manned Space Project with NOs. CMS-CSST-2021-A02. We thank Zhejie Ding, Yu Yu and Xiaodong Li for helpful discussions. We acknowledge the use of the High Performance Computing Resource in the Core Facility for Advanced Research Computing at the Shanghai Astronomical Observatory.

\section*{DATA AVAILABILITY}

Simulation data will be shared upon reasonable request. The data
supporting the plots within this article are available on reasonable
request to the corresponding author.

\bibliographystyle{mnras}
\bibliography{main}

\end{document}